\documentclass[12pt]{article}
\usepackage{graphicx,amsmath}
\usepackage{psfrag,graphicx,longtable}
\usepackage{setspace}

\begin{document} 
\singlespacing

   \title{Spectroscopy of high-energy states of lanthanide ions}
   \date{9 June 2010}

\maketitle

\begin{center}
   {Michael F. Reid}$^a$, 
   {Liusen Hu}$^{b,a}$,
   {Sebastian Frank}$^{a,c}$, \\
   {Chang-Kui Duan}$^{d}$, 
   {Shangda Xia}$^b$, and
   {Min Yin}$^b$
\end{center}

\bigskip

\noindent
  $^a${Department of Physics and Astronomy and
     MacDiarmid Institute for Advanced Materials and Nanotechnology, 
     University of Canterbury, Christchurch, New Zealand}\\
    Email: \texttt{mike.reid@canterbury.ac.nz}\\
  $^b${Department of Physics, University of Science and
    Technology of  China, Hefei 230026, China}\\
  $^c${Present Address: Department of Theoretical Physics, 
         University of Bayreuth, 95440 Bayreuth, Germany }\\
  $^d${Institute of Modern Physics, Chongqing University of
     Post and Telecommunications, Chongqing 400065, China}\\


\onehalfspacing

\begin{abstract}  
  We discuss recent progress and future prospects for the analysis of
  the 4f$^{N-1}$5d excited states of lanthanide ions in host materials.
  We demonstrate how {ab-initio} calculations for Ce$^{3+}$ in LiYF$_4$
  may be used to estimate crystal-field and spin-orbit parameters for
  the 4f$^1$ and 5d$^1$ configurations. We show how excited-state
  absorption may be used to probe the electronic and geometric structure
  of the 4f$^{N-1}$5d excited states in more detail and we illustrate
  the possibilities with calculations for Yb$^{2+}$ ions in SrCl$_2$.

\end{abstract}

\bigskip

\noindent {\it Keywords}: lanthanide, rare earth, crystal field, ab initio calculations,
 UV/vis spectroscopy, laser spectroscopy

\newpage

\section{Introduction}

The electronic structure of the 4f$^N$ configuration of lanthanide ions
in condensed matter environments has been a field of intense study since
the 1960s. The parametric models that were developed in the 1960s
\cite{Wyb65,Die68,CFR68} gained further sophistication through the 1970s
and 1980s (see, for example, Ref \cite{CGRR89}). For a recent review see
Ref. \cite{LiJa05}.

Many applications of lanthanide materials, such as scintillators,
lasers, and phosphors, involve the 4f$^{N-1}$5d excited configuration.
An understanding of these states, and other high-energy states, such as
charge-transfer, conduction-band, and exciton states
\cite{Do09,PeJoMc07,GrMa08,Do03a}, is crucial to the development of 
 better materials for such applications.

 Transitions to the 4f$^{N-1}$5d excited states of lanthanide ions have
 been studied for a long time, particularly for divalent ions
 \cite{McKi63,PiBrMc67} and for Ce$^{3+}$ \cite{ScWi68}, which have
 transitions in the UV region.  The availability of VUV synchrotron
 radiation from the 1970s allowed systematic experimental studies and
 some limited calculations to be carried out for the trivalent ions
 \cite{ElHeYe73,HeElYe75}. Good availability of VUV beamlines during the
 past two decades has led to renewed interest in these energy levels,
 and detailed analyses have been made over the past ten years
 \cite{RePiWeMe00,LaDoGiMaMo00,PiReWeSoMe02,PiReBuMe02}. These analyses
 have been reviewed in Ref. \cite{BuRe07}.

 Though the extension of the parametrized model is straightforward, its
 application to the 4f$^{N-1}$5d configuration is different from
 application to the 4f$^N$ configuration. The 4f$^N$ spectra consist of
 sharp lines, allowing detailed identification of energy levels and
 accurate fitting of parameters. In contrast, transitions between 4f$^N$
 and 4f$^{N-1}$5d configurations consist mainly of broad vibronic lines,
 so interpretation of the spectra is not so straightforward, and
 accurate determination of the parameters is difficult.  The
 crystal-field splittings of the 4f$^{N-1}$5d configuration are large
 compared to the splittings caused by the Coulomb interaction. Whereas
 for the 4f$^N$ configuration the crystal field can be considered as a
 minor perturbation to the free ion, this is not the case for the
 4f$^{N-1}$5d configuration, where the crystal field dominates the
 structure of the spectra.
 
 It is now possible to make reasonably accurate {ab-initio} calculations
 for the 4f$^N$ and 4f$^{N-1}$5d configurations of lanthanide ions in
 crystals
 \cite{OgWaToBr07,RuBaSe05,PaScBaSe06,SaSeBa09,YiZhXiYi05,AnKoDo07}.
 This raises the possibility of determining crystal-field and other
 parameters from such calculations. A method for doing this was
 developed in Ref. \cite{ReDuZh09}, and here is demonstrated for
 Ce$^{3+}$ ions in LiYF$_4$.

{Ab-initio} calculations can also give information about excited-state
bonding and geometry \cite{RuBaSe05,PaScBaSe06,SaSeBa09,AnKoDo07}. Some
excited states have shorter bond lengths than the ground state, and some
longer, consistent with various experiments. 

More information about the stucture and dynamics of the 4f$^{N-1}$5d
states may be obtained by using excited-state absorption (ESA). We
discuss how ESA might be used to reveal detailed electronic structure of
the 4f$^{N-1}$5d configuration.  Transitions where the 5d electron does
not change orbitals should give sharp-line spectra because there is no
change in bond length. On the other hand, transitions between states
where the 5d electron changes to a different orbital will again give
broad bands, with the width of the band a measure of the change in
excited-state bond length.

\section{Analysis of the 4f$^{N}$ and 4f$^{N-1}$5d 
configurations of\\  lanthanide ions in crystals }

Modelling of the 4f$^{N}$ and 4f$^{N-1}$5d configurations by
parametrized calculations has been reviewed in Ref. \cite{BuRe07}.
Such ``crystal field'' calculations make use of an ``effective
Hamiltonian'' \cite{HuFr93,BrRe98,LeReFaBu01} that acts solely within the
4f$^N$ and 4f$^{N-1}$5d configurations.  Rather than solving for the
eigenvalues and eigenstates of the full Hamiltonian, the effective
Hamiltonian is diagonalized within the model space (4f$^N$ and
4f$^{N-1}$5d configurations), and the expectation values of the
effective operators are evaluated between the model-space eigenvectors.

The effective Hamiltonian for the 4f$^N$ configuration may be written as
\cite{CGRR89,LiJa05}: 
\begin{eqnarray} \label{eq:hf}
H_\mathrm{f} &=& E_\mathrm{avg} 
  + \sum_{k=2,4,6} F^k f_k
  + \zeta_f A_\mathrm{so}
  + \alpha L(L+1)
  + \beta G(G_2)
  + \gamma G(R_7) 
\nonumber\\
  &+& \sum_{i=2,3,5,6,7,8} T^i t_i 
  + \sum_{h=0,2,4} M^k m_k
  + \sum_{k=2,4,6} P^k p_k
  + \sum_{k=2,4,6}\sum_{q} B^k_q C^{(k)}_q 
. 
\end{eqnarray}
$E_\mathrm{avg} $ is the energy difference between the ground-state energy and
the configuration center of gravity and is included to allow the
ground-state energy to be set to zero. The Coulomb interaction between
the 4f electrons is parametrized by the radial electronic integrals
$F^k$ and the spin-orbit interaction by $\zeta_f$. These parameters are
multiplied by appropriate tensor operators. The $\alpha$, $\beta$,
$\gamma$, $T^i$, $M^k$, and $P^k$ parameters represent higher-order
Coulomb and magnetic interactions. The major features of the spectra are
determined by the $F^k$ and $\zeta_f$ parameters, with the other
parameters giving small, but crucial, adjustments.

The $B^k_q$ are ``crystal-field'' parameters, which represent the
interaction of the 4f electrons with the surrounding ions. For 4f
electrons the non-zero parameters have $k = 2$, 4, 6, and $q$ values
determined by symmetry \cite{Wyb65,Die68,LiJa05}. 

For lanthanide 4f$^N$ configurations the Coulomb interaction gives
splittings of tens of thousands of cm$^{-1}$, the spin-orbit interaction
splittings of a few thousand cm$^{-1}$, and the crystal-field
interaction splittings of a few hundred cm$^{-1}$. Thus, the positions
of the electronic multiplets are quite similar in different crystals,
leading to the useful concept of a ``Dieke diagram'', which summarises
the energy levels of the entire lanthanide series \cite{Die68,CGRR89,LiJa05}

Transitions between $4f^N$ states are parity forbidden, so the
transitions are rather weak, consisting of magnetic-dipole and
``forced'' electric-dipole transitions. The latter are parametrized
according to the ``Judd-Ofelt'' approach \cite{Jud62,Ofe62,Re05}.

To extend the model to the 4f$^{N-1}$5d configuration we must add more
terms to our Hamiltonian. The extra terms are: 
\begin{eqnarray} \label{eq:hd}
H_\mathrm{d} &=& \Delta_E(\mathrm{fd}) 
  + \sum_{k=2,4} F^k(\mathrm{fd}) f_k
  + \sum_{k=1,3,5} G^k(\mathrm{fd}) g_k
  + \zeta (\mathrm{d}) A_\mathrm{so}(\mathrm{d})
\nonumber\\
  &+& \sum_{k=2,4}\sum_q  B^k_q(\mathrm{d}) C^{(k)}_q (\mathrm{d})
. 
\end{eqnarray}
In this equation $\Delta_E(\mathrm{fd})$ represents the energy
difference between the 4f$^{N}$ and 4f$^{N-1}$5d configurations, the
$F^k(\mathrm{fd})$ and $G^k(\mathrm{fd})$ direct and exchange Coulomb
interactions between the 4f and 5d electrons, and $B^k_q(\mathrm{d})$
the crystal-field interaction of the 5d electron with the surrounding
ions. This interaction is generally about twenty times larger than the
interaction for the 4f electrons, so the splittings due to this
interaction are tens of thousands of cm$^{-1}$, the same magnitude as
the Coulomb interaction.

This difference in interaction strength of the 4f and 5d electrons with
the ligands results in a shift in the excited-state geometry, and hence
the broad bands seen in transitions between the 4f$^{N}$ and
4f$^{N-1}$5d configurations.  Since transitions between the 4f$^{N}$ and
4f$^{N-1}$5d configurations generally consist of broad bands, rather
than sharp lines, analyses have made use of a variety of ways to
reducing the number of parameters, such as by using atomic calculation
for the ``free ion'' parameters. Examination of spectra for ions across
the series is also helpful, since certain features may be used to
calibrate the model. For example, the splitting between spin-allowed and
spin-forbidden transitions in the second half of the series may be used
to estimate the parameters for the Coulomb interaction between the 4f and
5d electrons \cite{PiReBuMe02}.

Calculations based on the above Hamiltonians give a 4f$^{N}\rightarrow$
4f$^{N-1}$5d spectrum with hundreds or thousands of transitions. But
since many of the details are obscured by the vibronic bands comparison
with experiment is not straightforward.  Therefore, it is helpful to
have some simple ways to catalog some of the experimental results.
Dorenbos \cite{Do00a,Do03,Do09} has given simple formulae to relate the
lowest-energy 4f$^{N}$ to 4f$^{N-1}$5d transition energies, and also
charge transfer transition energies, across the lanthanide series. In
another approach, Duan and co-workers \cite{DuXiReRu05,XiDuDeRu05} have
used a ``simplified model'', where the Coulomb interaction is
approximated by a simple exchange potential. Such models are helpful to
rationalizing the important features of the spectra.

There have been some attempts to treat the excited state vibrations in
more detail. Perturbations of the energies of the excited states by
electron-phonon coupling, which has been discussed in detail by Malkin
and co-workers \cite{MaSoMaSa07}. In some cases the individual
vibrations can be resolved, in which case the spectra may be analysed to
obtain excited state coordinate shifts. See, for example,
\cite{LiChEdReHu04,KaUrRe07}.

\section{Extracting parameters from {ab-initio} calculations}

First-principles ({ab-initio}) calculations of the electronic
structure of the 4f$^N$ and 4f$^{N-1}$5d configurations of lanthanide
ions in solids are now becoming common
\cite{RuBaSe05,YiZhXiYi05,OgWaToBr07,AnKoDo07}. Ishii, Ogasawara and
co-workers
\cite{IsToFuOgAd01,OgWaToIsBrIkTa05,WaIsFuOg06,OgWaToBr07,WaOg08} have
applied the DV-X$\alpha$ method for ions across the entire lanthanide
series.  Similar calculations have been performed by other workers
\cite{YiZhXiYi05,WaXiYi08}.  Seijo and co-workers
\cite{RuBaSe05,PaScBaSe06,SaSeBa09} have concentrated on a smaller
number of systems and have used a sophisticated quantum-mechanical
embedding scheme to  calculate realistic excited-state geometries and
potential-energy surfaces.

Though quite good agreement can now be obtained between {ab-initio}
calculations and experimental energies, it is desirable to compare the
{ab-initio} calculations with the parametric models. As discussed in the
previous section, parametric calculations are useful in part because
they give a way to compare spectra across the lanthanide series.
Parameters obtained in one ion may be extrapolated to others.  This may
be helpful in applying ab-initio calculations. For example,
crystal-field parameters may be calculated for Ce$^{3+}$, where there is
only one valence electron, and the parameters extrapolated to ions with
more complex electronic structure for which {ab-inito} calculations are
much more time consuming.

Comparisons at the level of parameters could also provide more detailed
tests of the {ab-initio} calculations. For example, it may be easier to
deduce which effects are being poorly represented in the {ab-initio}
calculations by extrating parameters from the calcuations. For example,
if correlation is poorly represented then the correlation parameters in
Eq.\ (\ref{eq:hf}) will be calculated to be too small.

In some cases, particularly in high symmetries, such as O$_\mathrm{h}$,
the parameters may be determined from {ab-initio} calculations by
fitting them to the calculated energy levels \cite{DuReXi07}.  However,
this is not always possible in low symmetry, where, for Ce$^{3+}$ there
are often more free parameters than energy levels. However, it is
possible to determine the 4f$^N$ or 4f$^{N-1}$5d ``effective
Hamiltonian'' \cite{HuFr93}, and therefore the parameters, directly, if
one has the eigenvectors, as well as the energies, for the relevant
states.  A straight-forward way of doing this has been discussed
recently \cite{ReDuZh09}.

We illustrate this method for the crystal-field and spin-orbit
parameters for the 4f$^1$ and 5d$^1$ configurations of Ce$^{3+}$ in
LiYF$_4$. Further details and calculations for other hosts will be
presented elsewhere. For these calculations we use the rather
unsophisticated DV-X$_\alpha$ method with a simple Madelung embedding.
This has many shortcomings, as discussed in Refs.\
\cite{RuBaSe05,PaScBaSe06,SaSeBa09}. However, our purpose here is to
illustrate the method, and this code is quite suitable for that. In
particular, it is easy to extract the eigenvectors from the calculation.

The DV-X$_\alpha$ technique was originally developed by Ellis and
co-workers for quantum chemical calculations of electronic and
structural properties of molecular systems \cite{RoEl75}.  It was
further developed by Adachi and co-workers \cite{AdTsSa78} and it has
been used for various calculations on lanthanide systems
\cite{IsToFuOgAd01,OgWaToIsBrIkTa05,WaIsFuOg06,WaOg08,YiZhXiYi05,WaXiYi08}.
These calculations are fully relativistic and so automatically include
the spin-orbit interaction. 

The coordinates for LiYF$_4$ were taken from Ref.  \cite{GaRy93}. For
the calculations reported here we used a small (CeF$_8$)$^{5-}$ cluster,
as used in Refs. \cite{OgWaToIsBrIkTa05,WaIsFuOg06,WaOg08}.

The calculated energy levels are presented in Table \ref{tab:energies}.
Also shown are the experimental 5d$^1$ energies and the results of
similar calculations by Ogasawara and co-workers
\cite{WaIsFuOg06,WaOg08}.  Since the main focus of this work is the
crystal-field splitting of the 4f$^1$ and 5d$^1$ configurations the
average energy of the 5d$^1$ configuration in our calculation 
has been shifted to match the
experimental average. The average before this adjustment is given in the
last line of the Table.

For Ce$^{3+}$ there is only one valence electron so the Hamiltonian only
includes the crystal-field and spin-orbit interactions.  In LiYF$_4$ the
energy levels consist of seven and five Kramer's doublets for 4f$^1$ and
5d$^1$ respectively. Taking into account the S$_4$ site symmetry, the
Hamiltonian may be written as:
\begin{equation}\label{eq:celiyf4}
H_\mathrm{Ce} = E_\mathrm{avg}
  +  \zeta(\mathrm{f}) 
  + \sum_{k=2,4,6} \sum_{q=0,4} B^k_q(\mathrm{f}) C^{(k)}_q 
  + \Delta_E(\mathrm{fd})  A_\mathrm{so}(\mathrm{d})
  + \zeta(\mathrm{d}) 
  + \sum_{k=2,4} \sum_{q=0,4} B^k_q(\mathrm{d}) C^{(k)}_q 
  . 
\end{equation}
In S$_4$ symmetry the $q=4$ parameters are complex.  However, we may
rotate about the $z$ axis to make one of our calculated parameters real
(e.g. see \cite{Wyb65,BuRe04}).  If we perform the rotation to make
$B^4_4$(f) real we find that in practise the imaginary parts of the
other parameters are small. Since the experimental parameters are
derived assuming a higher D$_{2d}$ symmetry, where all parameters are
real, we report real $B^k_4$ parameters by calculating the magnitude and
using the sign of the real part.

To use the approach of Ref.  \cite{ReDuZh09} we must identify 4f$^1$ and
5d$^1$ energy levels and eigenvectors. This is straightforward for the
small cluster used here.   In Table
\ref{tab:parameters} we list the crystal-field parameters for 4f$^1$ and
5d$^1$. We list experimental parameters for Pr$^{3+}$ and Nd$^{3+}$, as
well as Ce$^{3+}$, since the crystal-field parameters for Ce$^{3+}$ are
difficult to determine because there are the same number of parameters
as energy levels.

The calculated crystal-field parameters and spin-orbit parameters agree
quite well with experiment. However, as mentioned above, the
DV-X$_\alpha$ method has many shortcomings and more sophisticated
calculations would be required to obtain accurate results.  We note that
we could not have determined crystal-field parameters from only the
calculated energies since there are as many parameters as energies.
Thus, knowledge of the wave-functions are crucial to such a calculation.


For the small cluster the 4f and 5d orbitals can only mix with F$^-$
orbitals.  When we attempted calculations with a larger cluster the
shortcomings of the method became apparent and there was considerable
mixing of 5d orbitals with Y$^{3+}$ orbitals.  This has been previously
noted in Ref.\ \cite{IsToFuOgAd01}. While interactions with the
conduction band has an important effect on the high-energy states, as is
well-known from the lack of structure observed in the spectra and from
direct measurements of photoconductivity \cite{PeJoMc07}, the mixing we
obtained was too large to be physical.

\section{Probing electronic and geometrical structure with \\
excited-state  absorption}

The {ab-initio} calculations by Seijo and co-workers
\cite{RuBaSe05,PaScBaSe06,SaSeBa09} have challenged some common
assumptions. It is often assumed that the bond lengths always increase
when a lanthanide ion is excited from 4f$^N$ to 4f$^{N-1}$5d.  However,
the recent first-principles calculations cited above suggest that for
the lowest-energy 4f$^{N-1}$5d states the bond lengths are
\emph{shorter} than for the 4f$^N$ configuration. Only the magnitude,
not the sign, of this change in bond length may be easily estimated from
ground-state aborption (GSA) spectra.  Various evidence, such as
pressure measurements \cite{VaRoGoGuMaNaSaKr09} and EXAFS experiments in
the excited state \cite{BaEdRuRuSe05}, may be used to confirm these
calculations

These calculations also suggest that useful information may be obtained
by performing excited-state absorption (ESA) measurements between the
4f$^{N-1}$5d states. We now give a sketch of how such measurements might
be carried out.

For ions with $N>7$ the lowest 4f$^{N-1}$5d states have a higher spin
than the ground 4f$^N$ states and the lowest states with the same spin
as the ground state lie a few thousand cm$^{-1}$ higher.  Luminescence
is observed in many cases from both the ``High Spin'' (HS) and the
higher-energy ``Low Spin'' (LS) states of trivalent ions has been
observed \cite{PiReBuMe02}. Observation of Emission from more than one
excited state is particularly common for divalent ions (e.g.
\cite{PaDuTa08,BeGrGeGu06}.

In these cases it is possible to perform a very sensitive ESA experiment
by first populating the lower 4f$^{N-1}$5d states.  ESA may then be
obeserved by using a second excitation source by monitoring the
higher-energy emission. Such ``upconversion'' has already been observed
in Tm$^{2+}$ systems \cite{BeGrGeGu06}. However, those experiments only
demonstrated the effect by exciting at a single frequency. An ESA
spectrum was not reported.

The ab-initio calculations by Seijo and co-workers, such as
\cite{SaSeBa09}, suggest that for some excited-state transtions the
bond-length will not change because the 5d electron stays in the same
orbital. These transitions should be sharp, whereas those for which the
5d electron changes to a different orbital will be broad. If the bond
lengths when the 5d electron is in the lowest 5d orbitals are more
contracted than when it is in a 4f orbital and the bond lengths when the
electron is in a higher 5d orbital are more expanded, as indicated by
the calculations of Seijo and co-workers
\cite{RuBaSe05,PaScBaSe06,SaSeBa09}, then some of the transitions within
the 4f$^{N-1}$5d configuration will have even broader vibronic bands
than the 4f$^{N}$ to 4f$^{N-1}$5d transitions. 

These observations suggest that signficantly more information may be
obtained from ESA experiments than from conventional one-photon
experiments. The observation of sharp transitions within the
4f$^{N-1}$5d configuration (rather than the broad bands observed with
ground-state absopption) will allow a much more detailed analysis of the
elctronic structure of the excited states.  Detailed analyis of the
vibrational spectrum for transitions that do change bond length, such as
in Refs.\cite{LiChEdReHu04,KaUrRe07} will give information about changes
in excited state geometry.

We illustrate the information that might be obtained from ESA
measurements by considering the case of Yb$^{2+}$ in SrCl$_2$. This
system has been analysed in detail by Piper et al. \cite{PiBrMc67}, and
more recently by Pan et al. \cite{PaDuTa08}. These calculations make use
of our simple crystal-field model. Much more accurate simulations based
on ab-initio calculations such as Ref.\ \cite{SaSeBa09} should be
possible, but our point here is to illustrate the issues. 

\begin{figure}[htp]
\includegraphics[width = 17cm]{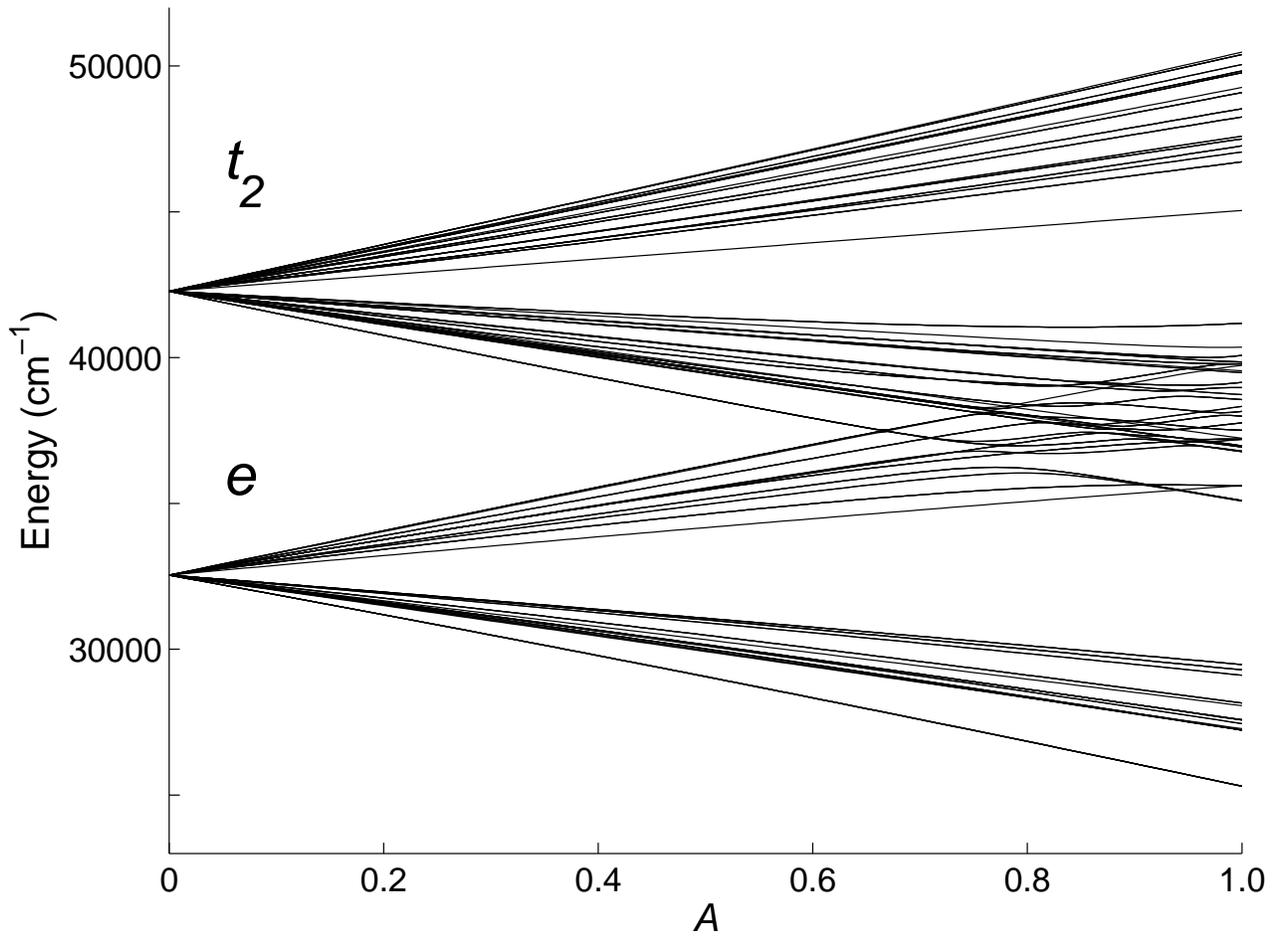}
\caption{
\label{fig:splitplot}
Calculated energy levels for Yb$^{2+}$ in SrCl$_2$. The parameter $A$
represents all of the 4f$^{13}$5d Hamiltonian apart from the 5d crystal
field (see text).  When $A=1$ the energies reproduce the calculation of
Ref. \cite{PaDuTa08}.
}
\end{figure}

In Figure \ref{fig:splitplot} we show the effect of ``switching on'' the
Hamiltonian, apart from the 5d crystal field. In O$_h$ symmetry there is
only one independent 5d crystal-field parameter, $B^4_0(\mathrm{d})$,
and the Hamiltonian may be written as:
\begin{equation}\label{eq:yb}
H_\mathrm{Yb} = 
 B^4_0(\mathrm{d}) \left[C^{(4)}_0 + \sqrt{\frac{5}{14}} 
  \left( C^{(4)}_4 + C^{(4)}_{-4} \right) \right] + A H_\mathrm{atomic}, 
\end{equation}
where $H_\mathrm{atomic}$ contains all other interactions apart from the
5d crystal field (and thus is not truly ``atomic'', since it contains
the 4f crystal field). In Figure \ref{fig:splitplot} the parameter $A$
is varied from 0 to 1. When $A=0$ we have only the 5d crystal-field
splitting, with the lowest states having the 5d electron in an $e$
orbital and the higher states having the 5d electron in a $t_2$ orbital.
When $A=1$ we reproduce the calculation of Ref. \cite{PaDuTa08}.

Though the lowest 4f$^{13}$5d state has the 5d electron in an $e$
orbital, it is clear that some states for which the d electron is in an
$e$ or a $t_2$ orbital overlap and mix when $A=1$. We may calculate this
mixing by expressing the eigenvectors for $A=1$ as linear combinations
of the eigenvectors for $A=0$. This allows us to simulate the ESA
spectrum by assigning appropriate line-widths for the various
transitions.

\begin{figure}[htp]
\includegraphics[width = 17cm]{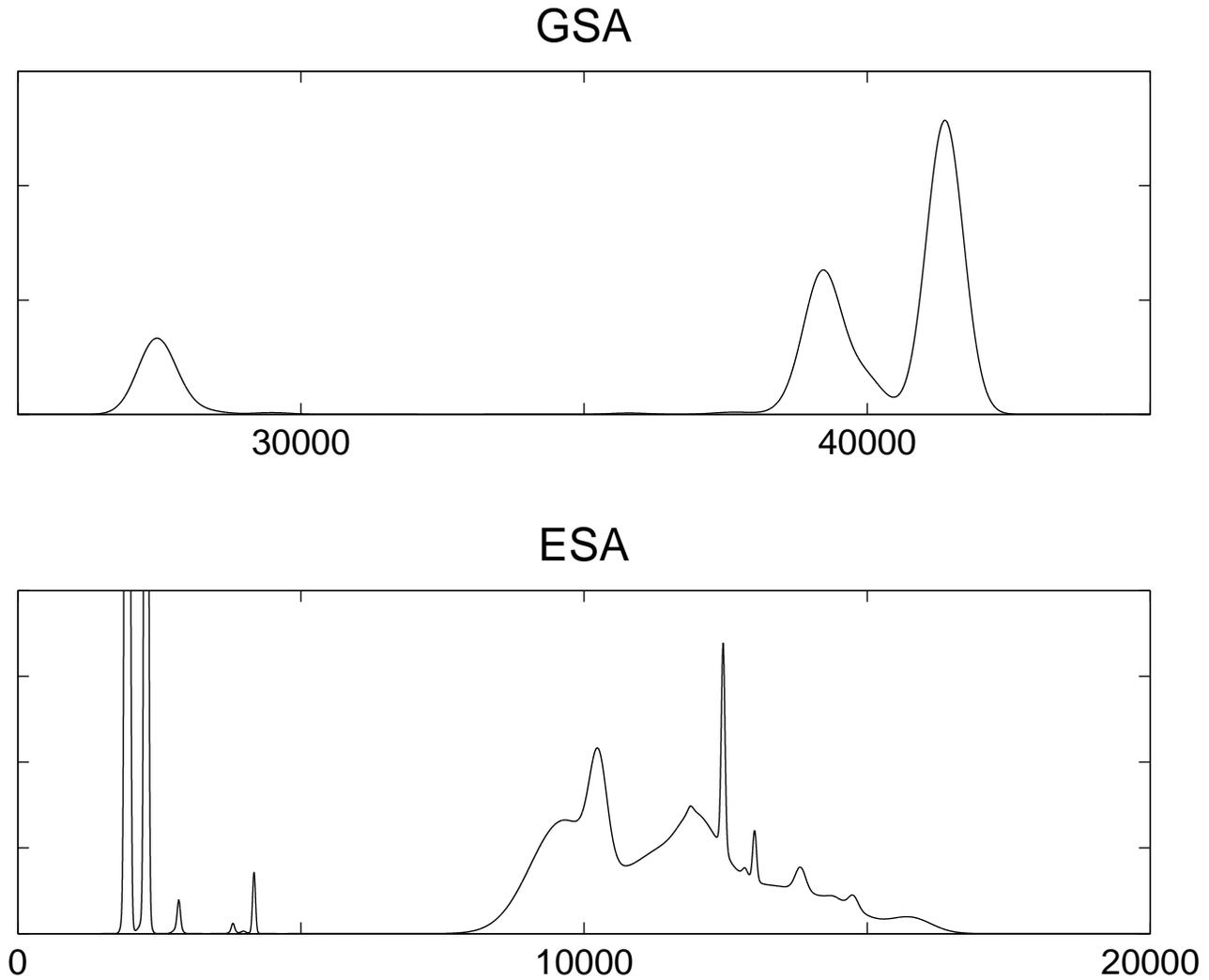}
\caption{
\label{fig:fdAbsESA}
Simulated ground-state absorption (GSA) and excited-state absorption
(ESA) spectra for Yb$^{2+}$ in SrCl$_2$. The horizontal axes have units
of cm$^{-1}$.  The ESA is shifted so that the final states are at the
same position on both graphs.  Note that the transitions from the ground
state to the the lowest 4f$^{13}$5d state at approximately 25000
cm$^{-1}$ is forbidden.  Linewidths for the GSA transitions are 650
cm$^{-1}$, linewidths for the ESA transitions range from 50 cm$^{-1}$ to
1300 cm$^{-1}$, depending on the fraction of 5d:$t_2$ in the final
state.  }
\end{figure}

In Figure \ref{fig:fdAbsESA} we show calculated GSA and ESA
spectra. The GSA transitions are assigned a vibronic linewidth of 650
cm$^{-1}$. We assume, for simplicity, that the 5d:$e$ states are
contracted and the 5d:$t_2$ states expanded by the same amount relative
to the 4f$^{14}$ ground state. Therefore, we assign a small linewidth
(50 cm$^{-1}$) to ESA transitions to pure 5d:$e$ states and a linewidth
\emph{twice} the GSA linewidth to ESA transitions to pure 5d:$t_2$
states. Mixed states have intermediate linewidths according to there
proportion of 5d:$t_2$. In the O$_h$ site symmetry of Yb$^{2+}$ in
SrCl$_2$ the ESA transitions within the 4f$^{13}$5d configuration are
only magnetic-dipole allowed, and so their dipole strengths are about
1000 times smaller than the electric-dipole allowed GSA transitions.
However, they are still strong compared to typical 4f$^N$ transitions,
and we note that ESA experiments similar to those discussed here have
been carried out for transitions within the 4f$^{7}$ configuration of
Gd$^{3+}$ \cite{PeVeScMeReBu05}.

The transition to the HS state is forbidden, so does not appear in the
simulated GSA spectrum at about 25,000 cm$^{-1}$. It is clear from
Figure \ref{fig:fdAbsESA} that an ESA spectrum could give much more
information than the GSA spectrum.

Yb$^{2+}$ in SrCl$_2$ has drawbacks for this sort of experiment.  It is
difficult to populate the lowest excited state due to the transition
from the ground state to the lowest excited state being electric-dipole
and magnetic-dipole forbidden due to point-group selection rules and the
non-radiative relaxation from higher states being very slow
\cite{PaDuTa08}. More promising candidates are Tm$^{2+}$ systems, for
which up-conversion has already been observed \cite{BeGrGeGu06} and
experiments on these systems are currently underway.

\section{Conclusions}

It is clear that there are many interesting aspects of excited states of
lanthanide ions remaining to be explored. We have demonstrated that it
is possible to calculate crystal-field parameters from {ab-initio}
calculations and have suggested excited-state absorption experiments
that should allow the exploration of excited states involving 5d
electrons in much more detail than in the past.

\section*{Acknowledgements}

M.F.R. acknowledges travel support from the MacDiarmid Institute and
support from the International Conference on f-Elements. \\
L.H. acknowledges support from a Chinese Government
Scholarship for his visit to New Zealand.\\
C.-K.D acknowledges National Science Foundation of China Grant No.\ 
10874173 for financial support.

\newpage


\newpage
\begin{table}
  \caption{Experimental and calculated energy levels for the 4f$^1$ and 5d$^1$ 
    configurations of Ce$^{3+}$
    in LiYF$_4$. 
    \label{tab:energies}}
\begin{tabular}{lrrr}
\hline
              &Experiment \cite{PiWeMeRe01}   &{Calculation (this work)}
                                                     &Calculation(Ref. \cite{WaIsFuOg06}) \\
\hline
4f$^1$        &0             &0                         &0         \\ 
              &              &616                       &129       \\ 
              &              &672                       &492       \\ 
              &              &2604                      &2807      \\ 
              &              &2971                      &2896      \\ 
              &              &3316                      &3041      \\ 
              &              &3461                      &3646      \\ 
\\         
5d$^1$        &33400         &33764                     &40086     \\ 
              &41200         &43078                     &45974     \\ 
              &48600         &48145                     &51781     \\ 
              &50500         &48739                     &52829     \\ 
              &53000         &52662                     &56781     \\ 
\\         
5d average    &45277         &30864                     &49490     \\ 
\hline
\end{tabular}

\end{table}

\vfill

~ 

\clearpage

\begin{table}
  \caption{Crystal-field and spin-orbit parameters  for the 4f$^1$ and
    5d$^1$ configurations of Ce$^{3+}$  in LiYF$_4$. 
    \label{tab:parameters}}
\begin{tabular}{lrrrrr}
\hline
       Parameter &{Calculation}    &\multicolumn{3}{c}{Experiment}\\
                 &                  &Ce$^{3+}$$^a$&Pr$^{3+}$ $^b$  &Nd$^{3+}$$^c$ \\
\hline                                                               
       B$^2_0$(f)  &1228            &481         &489          &409   \\
       B$^4_0$(f)  &$-$538          &$-$1150     &$-$1043      &$-$1135    \\  
       B$^4_4$(f)  &$-$1001         &$-$1228     &$-$1242      &$-$1216    \\
       B$^6_0$(f)  &12              &$-$89       &$-$42        &27    \\
       B$^6_4$(f)  &$-$723          &$-$1213     &$-$1213      &$-$1083    \\
       $\zeta$(f)  &748             &615         &731          &871    \\
\\                                                      
       B$^2_0$(d)  &5338            &4673        &7290         &    \\
       B$^4_0$(d)  &$-$12155        &$-$18649    &$-$14900     &    \\
       B$^4_4$(d)  &$-$23448        &$-$23871    &17743        &    \\
       $\zeta$(d)  &894             &1082        &906          &    \\
\hline  
\end{tabular}

$^a$ Extrapolated and fitted parameters from Ref. \cite{PiWeMeRe01}.\\
$^b$ Fitted parameters from Ref. \cite{LaDoGiMaMo00}.\\
$^c$ Fitted parameters from Ref. \cite{GB96}\\

\end{table}

\vfill 

~



\clearpage

\clearpage

\end{document}